# Relating the solutions of the Dirac equation with a background electric potential to solutions with a background pseudoscalar potential.

Dan Solomon
Rauland-Borg Corporation
1802 W. Central Road
Mount Prospect, IL 60056 USA
Email: dan.solomon@rauland.com
Oct. 19, 2012

**Abstract.**

We compare two different solutions of the Dirac equation in (1+1) dimensions. One solution is for a fermion in the presence of an electric potential and the other is for a fermion in the presence of a pseudoscalar potential. It is shown that for properly defined potentials one can easily relate the solutions of one system to the solutions of the other. In effect, solving one problem gives the solution to both. In addition, the vacuum charge density is calculated in the both cases and it is shown how this result is impacted by the presence of anomalies in quantum field theory.

## 1. Introduction.

In this paper we will consider the Dirac equation in (1+1) dimensions in the presence of two different external potentials. One potential will be a static electric potential and the other potential will be a pseudoscalar potential. We will show that for properly defined potentials we can readily relate the solutions of one potential to the solutions of the other.

Consider the Dirac equation,

$$i\frac{\partial \varphi(z,t)}{\partial t} = H(z,t)\varphi(z,t) \qquad (1.1)$$

where $\varphi(z,t)$ is the wave function. We will consider two different Hamiltonians, $H_A(z,t)$ and $H_B(z,t)$, corresponding to System A and System B, respectively. System A will describe a fermion in the presence of a static electric potential. For this case the Hamiltonian is given as,

$$H_A(z,t) = -i\gamma^5 \frac{\partial}{\partial z} + m\gamma^0 + V(z) \qquad (1.2)$$

where $m$ is the mass, $V(z)$ is an applied static electric potential, and $\gamma^5 = \gamma^0 \gamma^1$ with $\gamma^0$ and $\gamma^1$ being the gamma matrices in two dimensions. They satisfy,

$$\gamma^0 \gamma^1 = -\gamma^1 \gamma^0; \quad \gamma^0 \gamma^0 = 1; \quad \gamma^1 \gamma^1 = -1; \quad \gamma^{0\dagger} = \gamma^0; \quad \gamma^{1\dagger} = -\gamma^1 \qquad (1.3)$$

An example of a solution to this equation for square well potential is given by these references [1,2].

The System B Hamiltonian is given by,

$$H_B(z,t) = -i\gamma^5 \frac{\partial}{\partial z} + m\gamma^0 e^{i\gamma^5 \theta(z)} \qquad (1.4)$$

This describes a fermion in the presence of a psuedoscalar potential. Some system B type solutions to the Dirac equation are discussed by these papers [2, 3,4,5].

Next assume that $\theta(z)$ is given in terms of $V(z)$ by the following expression,

$$\theta(z) = -2 \int_0^z V(z') dz' \qquad (1.5)$$

Given this relationship we will show that we can easily related System B solutions of the Dirac equation to the System A solutions. This possibility has been commented on briefly by others [3] however it is the purpose of this paper to discuss this problem in more detail and also to consider the effect of anomalies on the results.

## 2. A Solution to the Dirac Equation.

In this section we build on a result from Ref. [6] and determine the solution to Eq. (1.1) given the following Hamiltonian,

$$H(z,t) = -i\gamma^5 \frac{\partial}{\partial z} + M(z,t) \gamma^0 e^{i\gamma^5 \theta(z,t)} + (V(z,t) + U(z,t)) \qquad (2.1)$$

In the above assume that,

$$U(z,t) = \frac{\partial f(z,t)}{\partial t} + \frac{1}{2} \frac{\partial \theta(z,t)}{\partial z} \quad \text{where} \quad \frac{\partial f(z,t)}{\partial z} = -\frac{1}{2} \frac{\partial \theta(z,t)}{\partial t} \qquad (2.2)$$

It is shown in Appendix 1 that the solution to Eq. (1.1) given (2.1) and (2.2) is,

$$\varphi(z,t) = e^{-if(z,t)} e^{-i\gamma^5 \theta(z,t)/2} \chi(z,t) \qquad (2.3)$$

where $\chi(z,t)$ is the solution to,

$$i\frac{\partial \chi(z,t)}{\partial t} = \left(-i\gamma^5 \frac{\partial}{\partial z} + M(z,t)\gamma^0 + V(z,t)\right)\chi(z,t) \tag{2.4}$$

We will now use these results to show that System A Hamiltonian, $H_A$, can evolve into the System B Hamiltonian, $H_B$, by application of the appropriate field.

Refer to (2.1) and let $M(z,t) = m$ and $V(z,t) = V(z)$ and assume that $U(z,t)$ along with $f(z,t)$ is defined by (2.2). In addition to this let,

$$\theta(z,t) = \lambda(t)\theta(z); \quad \theta(z) = -2\int_0^z V(z')dz' \tag{2.5}$$

where the function $\lambda(t)$ is used to turn on the field. $\lambda(t)$ changes from an initial value of zero to a final value of one per the following,

$$\lambda(t) = 0 \text{ for } t < 0; \quad \lambda(t) \neq 0 \text{ for } 0 \leq t \leq t_f; \quad \lambda(t) = 1 \text{ for } t > t_f \tag{2.6}$$

From the above relationships we find that for $t < 0$, $H(z, t<0) = H_A(z)$ and for $t > t_f$, $H(z, t>t_f) = H_B(z)$. That is, under the action of the field given in (2.5) along with (2.2) the Hamiltonian operator changes from $H_A$ to $H_B$.

Using these conditions in Eq. (2.1) through (2.4) the wave function is given by,

$$\varphi(z,t) = e^{-if(z,t)} e^{-i\gamma^5 \lambda(t)\theta(z)/2} \varphi_A(z,t) \tag{2.7}$$

where $\varphi_A(z,t)$ is the solution to,

$$i\frac{\partial \varphi_A(z,t)}{\partial t} = H_A(z)\varphi_A(z,t) \tag{2.8}$$

where $H_A(z)$ is the is the System A Hamiltonian which was defined by Eq. (1.2). When $t > t_f$ the quantity $f(z,t) = 0$ since $\theta(z, t>t_f) = \theta(z)$ is time independent. Therefore if $t > t_f$,

$$\varphi_B(z,t) = e^{-i\gamma^5 \theta(z)/2} \varphi_A(z,t) \tag{2.9}$$

which satisfies the equation,

$$i\frac{\partial \varphi_B(z,t)}{\partial t} = H_B(z)\varphi_B(z,t) \tag{2.10}$$

where $H_B$ is the System B Hamiltonian given by (1.4). Therefore the application of the field (2.5) along with (2.2) causes the wave function to evolve from System A to System B where the System A and B wave functions are related by (2.9).

## 3. The field operator and mode solutions.

In quantum field theory the field operator $\hat{\psi}(z,t)$ can be expanded in terms of the mode solutions of the Hamiltonian. For the System A Hamiltonian these mode solutions are given by,

$$u_{A,j,n}(z,t) = u_{A,j,n}(z)e^{-iE_{j,n}t} \tag{3.1}$$

They obey the following eigenvalue equation which is derived from Eq. (2.8),

$$E_{j,n}u_{A,j,n}(z) = H_A(z)u_{A,j,n}(z) \tag{3.2}$$

where $j = +1$ for positive energy states and $j = -1$ for negative energy states. In general these solutions include bound states and continuum states. The System A field operator is then,

$$\hat{\psi}_A(z,t) = \sum_n \left( \hat{b}_n u_{A,+1,n}(z)e^{-iE_{+1,n}t} + \hat{d}_n^\dagger u_{A,-1,n}(z)e^{-iE_{-1,n}t} \right) \tag{3.3}$$

where $\hat{b}_p\left(\hat{b}_p^\dagger\right)$ are the destruction (creation) operator for fermions and $\hat{d}_p\left(\hat{d}_p^\dagger\right)$ are the destruction (creation) operators for anti-fermions. They obey the usual anti-commutation relationships,

$$\left\{\hat{b}_n, \hat{b}_{n'}^\dagger\right\} = \delta_{n'n}; \quad \left\{\hat{d}_n, \hat{d}_{n'}^\dagger\right\} = \delta_{n'n} \tag{3.4}$$

with all other anti-commutations equal to zero. The vacuum state $|\Omega_0\rangle$ satisfies $\hat{b}_n|\Omega_0\rangle = 0$ and $\hat{d}_n|\Omega_0\rangle = 0$.

Under the action of the applied field (2.5), (2.2), and (1.5) each mode evolves in time and, referring to (2.7), we obtain,

$$u_{j,n}(z,t) = e^{-if(z,t)}e^{-i\gamma^5\lambda(t)\theta(z)/2}u_{A,j,n}(z,t) \tag{3.5}$$

For $t > t_f$ the Hamiltonian has evolved into $H_B(z)$, and the System B mode solutions are given by,

$$u_{B,j,n}(z,t) = e^{-i\gamma^5\theta(z)/2}u_{A,j,n}(z,t) \tag{3.6}$$

These System B mode solutions obey the eigenvalue equation,

$$E_{j,n} u_{B,j,n}(z) = H_B(z) u_{B,j,n}(z) \tag{3.7}$$

and the System B field operator can be expressed as,

$$\hat{\psi}_B(z,t) = \sum_n \left( \hat{b}_n u_{B,+1,n}(z) e^{-iE_{+1,n}t} + \hat{d}_n^\dagger u_{B,-1,n}(z) e^{-iE_{-1,n}t} \right) \tag{3.8}$$

Referring to (3.6) and (3.3) this can also be written as,

$$\hat{\psi}_B(z,t) = e^{-i\gamma^5 \theta(z)/2} \hat{\psi}_A(z,t) \tag{3.9}$$

So the System A modes evolve directly over into System B modes. Continuum modes evolve into continuum modes and bound state modes evolve into bound state modes. The corresponding modes $u_{A,j,n}$ and $u_{B,j,n}$ have the same eigenvalue $E_{j,n}$ so that the energy of a mode does not change when the system is evolved from System A to System B.

## 4. Charge Density.

The charge density operator can be defined by $\hat{\rho}(z,t) = \left[ \hat{\psi}^\dagger(z,t), \hat{\psi}(z,t) \right]$. The System A charge density operator is, therefore,

$$\hat{\rho}_A(z,t) = \left[ \hat{\psi}_A^\dagger(z,t), \hat{\psi}_A(z,t) \right] \tag{4.1}$$

Use (3.3) in the above to obtain,

$$\langle \Omega_0 | \hat{\rho}_A(z,t) | \Omega_0 \rangle = \sum_n \left( u_{A,-1,n}^\dagger(z) u_{A,-1,n}(z) - u_{A,+1,n}^\dagger(z) u_{A,+1,n}(z) \right) \tag{4.2}$$

where $\langle \Omega_0 | \hat{\rho}_A(z,t) | \Omega_0 \rangle$ is the vacuum charge density.

The System B charge is,

$$\hat{\rho}_B(z,t) = \left[ \hat{\psi}_B^\dagger(z,t), \hat{\psi}_B(z,t) \right] \tag{4.3}$$

Since we are working in the Heisenberg picture the time dependence of the field is associated with the field operator while the state vector is constant in time. Therefore the vacuum state vector $|\Omega_0\rangle$ does not change as we evolve from System A to System B as a result of the application of the field.

Use (3.8) in the above to obtain,

$$\langle \Omega_0 | \hat{\rho}_B(z,t) | \Omega_0 \rangle = \sum_n \left( u_{B,-1,n}^\dagger(z) u_{B,-1,n}(z) - u_{B,+1,n}^\dagger(z) u_{B,+1,n}(z) \right) \tag{4.4}$$

This is the vacuum charge density in System B. From (3.6) we have
$u_{A,j,n}^\dagger(z)u_{A,j,n}(z) = u_{B,j,n}^\dagger(z)u_{B,j,n}(z)$. Use this in the above and refer to (4.2) to obtain,

$$\langle \Omega_0 | \hat{\rho}_A(z,t) | \Omega_0 \rangle = \langle \Omega_0 | \hat{\rho}_B(z,t) | \Omega_0 \rangle \tag{4.5}$$

The conclusion is that the System B vacuum charge density is equal to the System A vacuum charge density.

## 5. The effect of the anomaly.

There is a potential problem with result of the last section in that is does not take into account the fact that quantum field theory has anomalies. The existence of anomalies requires us to modify the definition of the charge density operator using point split regularization [2, 7, 8]. We will apply point splitting in the spatial dimension. For System A the spatially point split charge density operator can be written as,

$$\hat{\rho}_A(z,t;\varepsilon) = \frac{1}{2}\left[\hat{\psi}_A^\dagger(z+\varepsilon,t)\hat{\psi}_A(z,t) - \hat{\psi}_A(z+\varepsilon,t)\hat{\psi}_A^\dagger(z,t)\right] \tag{5.1}$$

where $\varepsilon \to 0$. Using the equal time anti-commutation relationship
$\{\hat{\psi}_\alpha(z',t), \hat{\psi}_\beta^\dagger(z,t)\} = \delta_{\alpha\beta}\delta(z-z')$ (where $\alpha$ and $\beta$ are spinnor indices) we can write this as,

$$\hat{\rho}_A(z,t;\varepsilon) = \frac{1}{2}\left[\hat{\psi}_A^\dagger(z+\varepsilon,t)\hat{\psi}_A(z,t) + \hat{\psi}_A^\dagger(z,t)\hat{\psi}_A(z+\varepsilon,t)\right] \tag{5.2}$$

To demonstrate the impact of point splitting we will consider System A where $V(z)$ is the square well potential given by,

$$V(z) = \begin{cases} -\eta & \text{for } |z| < a/2 \\ 0 & \text{for } |z| > a/2 \end{cases} \tag{5.3}$$

where $m \geq \eta \geq 0$. This system was examined in [1,2]. It was shown by Capri et al[2] (and further discussed in [9]) that the point split vacuum charge density for this potential is given by,

$$\langle \Omega_0 | \hat{\rho}_A(z,t;\varepsilon) | \Omega_0 \rangle = \langle \Omega_0 | \hat{\rho}_A(z,t) | \Omega_0 \rangle - \frac{V(z)}{\pi} \tag{5.4}$$

That is, the effect of point split regularization is to subtract the quantity $V(z)/\pi$ from the vacuum charge density $\langle \Omega_0 | \hat{\rho}_A(z,t) | \Omega_0 \rangle$ that is calculated without point splitting.

Referring to (1.5) and (5.3) the System B field is given by,

$$\theta(z) = 2\eta \times \begin{cases} a/2 & \text{for } z > a/2 \\ z & \text{for } |z| < a/2 \\ -a/2 & \text{for } z < a/2 \end{cases} \qquad (5.5)$$

The System B point split charge density operator is,

$$\hat{\rho}_B(z,t;\varepsilon) = \frac{1}{2}\left[\hat{\psi}_B^\dagger(z+\varepsilon,t)\hat{\psi}_B(z,t) + \hat{\psi}_B^\dagger(z,t)\hat{\psi}_B(z+\varepsilon,t)\right] \qquad (5.6)$$

Use (2.9) to obtain,

$$\hat{\psi}_B^\dagger(z+\varepsilon,t)\hat{\psi}_B(z,t) = \hat{\psi}_A^\dagger(z+\varepsilon,t)e^{+i\gamma^5\theta(z+\varepsilon)/2}e^{-i\gamma^5\theta(z)/2}\hat{\psi}_A(z,t) \qquad (5.7)$$

In the limit $\varepsilon \to 0$ we obtain,

$$\hat{\psi}_B^\dagger(z+\varepsilon,t)\hat{\psi}_B(z,t) = \hat{\psi}_A^\dagger(z+\varepsilon,t)e^{+\frac{i\gamma^5}{2}\frac{d\theta(z)}{dz}\varepsilon}\hat{\psi}_A(z,t) \qquad (5.8)$$

One again, using $\varepsilon \to 0$, we obtain,

$$\hat{\psi}_B^\dagger(z+\varepsilon,t)\hat{\psi}_B(z,t) = \hat{\psi}_A^\dagger(z+\varepsilon,t)\left(1 + \frac{i\gamma^5}{2}\frac{d\theta(z)}{dz}\varepsilon\right)\hat{\psi}_A(z,t) \qquad (5.9)$$

Rearrange terms to obtain,

$$\hat{\psi}_B^\dagger(z+\varepsilon,t)\hat{\psi}_B(z,t) = \hat{\psi}_A^\dagger(z+\varepsilon,t)\hat{\psi}_A(z,t) + \frac{i}{2}\hat{\psi}_A^\dagger(z+\varepsilon,t)\gamma^5\hat{\psi}_A(z,t)\frac{d\theta(z)}{dz}\varepsilon \qquad (5.10)$$

Similarly,

$$\hat{\psi}_B^\dagger(z,t)\hat{\psi}_B(z+\varepsilon,t) = \hat{\psi}_A^\dagger(z,t)\hat{\psi}_A(z+\varepsilon,t) - \frac{i}{2}\hat{\psi}_A^\dagger(z,t)\gamma^5\hat{\psi}_A(z+\varepsilon,t)\frac{d\theta(z)}{dz}\varepsilon \qquad (5.11)$$

Therefore,

$$\hat{\rho}_B(z,t;\varepsilon) = \hat{\rho}_A(z,t;\varepsilon) + \frac{i}{4}\begin{bmatrix}\hat{\psi}_A^\dagger(z+\varepsilon,t)\gamma^5\hat{\psi}_A(z,t) \\ -\hat{\psi}_A^\dagger(z,t)\gamma^5\hat{\psi}_A(z+\varepsilon,t)\end{bmatrix}\frac{d\theta(z)}{dz}\varepsilon \qquad (5.12)$$

It is shown in Appendix 2 that,

$$\frac{i}{4}\langle\Omega_0|\begin{bmatrix}\hat{\psi}_A^\dagger(z+\varepsilon,t)\gamma^5\hat{\psi}_A(z,t) \\ -\hat{\psi}_A^\dagger(z,t)\gamma^5\hat{\psi}_A(z+\varepsilon,t)\end{bmatrix}|\Omega_0\rangle = -\frac{1}{2\pi\varepsilon} \qquad (5.13)$$

From (1.5) $d\theta(z)/dz = -2V(z)$. Use this to obtain,

$$\langle\Omega_0|\hat{\rho}_B(z,t;\varepsilon)|\Omega_0\rangle = \langle\Omega_0|\hat{\rho}_A(z,t;\varepsilon)|\Omega_0\rangle + \frac{V(z)}{\pi} \qquad (5.14)$$

Therefore when the point split procedure is used the System B vacuum charge density does not equal the System A vacuum charge density. This is in contradiction to the results of the last section.

Use (5.4) and (4.5) in the above to obtain,

$$\langle\Omega_0|\hat{\rho}_B(z,t;\varepsilon)|\Omega_0\rangle = \langle\Omega_0|\hat{\rho}_A(z,t)|\Omega_0\rangle = \langle\Omega_0|\hat{\rho}_B(z,t)|\Omega_0\rangle \quad (5.15)$$

For System B the charge density as calculated by point splitting is equal to the charge density as calculated without point splitting.

## 6. Conclusion.

We have examined two systems with different potentials and shown that the solutions of one can be related to the solutions of the other. In addition, the vacuum charge density was determined for each system. If the charge density operator is defined without point splitting then it is shown that the vacuum charge density for both potentials is the same. However if we define the charge density operator using point splitting it is shown that the vacuum charge densities are related according to (5.14) and are not equal.

## Appendix 1.

We what to find a solution to the following equation using the conditions in (2.2),

$$i\frac{\partial\varphi(z,t)}{\partial t} = \left[-i\gamma^5\frac{\partial}{\partial z} + M(z,t)\gamma^0 e^{i\gamma^5\theta(z,t)} + (V(z,t)+U(z,t))\right]\varphi(z,t) \quad (7.1)$$

Let,

$$\varphi(z,t) = e^{-if(z,t)} e^{-i\gamma^5\theta(z,t)/2} \chi(z,t) \quad (7.2)$$

From this we obtain,

$$\frac{\partial\varphi}{\partial t} = -i\left(\frac{\partial f}{\partial t} + \frac{\gamma^5}{2}\frac{\partial\theta}{\partial t}\right)\varphi + e^{-if} e^{-i\gamma^5\theta/2}\frac{\partial\chi}{\partial t} \quad (7.3)$$

and,

$$\frac{\partial\varphi}{\partial z} = -i\left(\frac{\partial f}{\partial z} + \frac{\gamma^5}{2}\frac{\partial\theta}{\partial z}\right)\varphi + e^{-if} e^{-i\gamma^5\theta/2}\frac{\partial\chi}{\partial z} \quad (7.4)$$

Use the above relationships along with $\gamma^5\cdot\gamma^5 = 1$ and (2.2) to obtain,

$$ie^{-if} e^{-i\gamma^5\theta/2}\frac{\partial\chi}{\partial t} = -ie^{-if} e^{-i\gamma^5\theta/2}\gamma^5\frac{\partial\chi}{\partial z} + M\gamma^0 e^{i\gamma^5\theta}\varphi + V\varphi \quad (7.5)$$

Next we will multiply both sides by $e^{+if}e^{+i\gamma^5\theta/2}$ to obtain,

$$i\frac{\partial \chi}{\partial t} = -i\gamma^5 \frac{\partial \chi}{\partial z} + M\gamma^0 \chi + V\chi \tag{7.6}$$

which is (2.4) in the text. To obtain the above result we have used

$$e^{+i\gamma^5\theta/2}\gamma^0 = \gamma^0 e^{-i\gamma^5\theta/2}$$

## Appendix 2.

It was noted by Capri et al[2] that the short distance behavior of the two point function in the presence of an external potential approaches that of a free field as $\varepsilon \to 0$. Therefore we will evaluate,

$$\left\langle \Omega_0 \left| \begin{bmatrix} \hat{\psi}_A^\dagger(z+\varepsilon,t)\gamma^5\hat{\psi}_A(z,t) \\ -\hat{\psi}_A^\dagger(z,t)\gamma^5\hat{\psi}_A(z+\varepsilon,t) \end{bmatrix} \right| \Omega_0 \right\rangle$$

for the $V(z)=0$ in the limit $\varepsilon \to 0$ and use the result in (5.12).

We work in a representation $\gamma^0 = \sigma_3$ and $\gamma^5 = \sigma_1$ where the $\sigma_j$ are the Pauli matrices. For the free field case the mode solutions are given by,

$$E_{j,p}v_{j,p}(z) = \left(-i\sigma_1 \frac{\partial}{\partial z} + m\sigma_3\right)v_{j,p}(z); \quad v_{j,p}(z) = w_{j,p}e^{-ipz} \tag{8.1}$$

where,

$$w_{j,p}(z) = N_{j,p}\begin{pmatrix} 1 \\ p/(E_{j,p}+m) \end{pmatrix}; \quad N_{j,p} = \sqrt{\frac{E_{j,p}+m}{2E_{j,p}}}; \quad E_{j,p} = j\sqrt{p^2+m^2} \tag{8.2}$$

where $j=1$ for positive energy states and $j=-1$ for negative energy states.

From the above we obtain,

$$v_{j,p}^\dagger(z+\varepsilon)v_{j,p}(z) = \frac{p}{E_{j,p}}e^{-ip\varepsilon}; \quad v_{j,p}^\dagger(z+\varepsilon)v_{j,p}(z) = \frac{p}{E_{j,p}}e^{+ip\varepsilon} \tag{8.3}$$

This yields,

$$\left\langle 0 \left| \begin{bmatrix} \hat{\psi}^\dagger(z+\varepsilon,t)\gamma^5\hat{\psi}(z,t) \\ -\hat{\psi}^\dagger(z,t)\gamma^5\hat{\psi}(z+\varepsilon,t) \end{bmatrix} \right| 0 \right\rangle = \frac{1}{2\pi}\int_{-\infty}^{+\infty}\left\{\left(\frac{pe^{+ip\varepsilon}}{-\sqrt{p^2+m^2}}\right) - \left(\frac{pe^{-ip\varepsilon}}{-\sqrt{p^2+m^2}}\right)\right\}dp \tag{8.4}$$

In the limit $\varepsilon \to 0$ we can replace $\sqrt{p^2+m^2}$ with $|p|$. Next use the following relationship,

$$\int_0^{+\infty} e^{+ip\varepsilon} dp = -\frac{1}{i\varepsilon} \tag{8.5}$$

in (8.4) to obtain,

$$\langle 0 | \begin{bmatrix} \hat{\psi}^\dagger(z+\varepsilon,t)\gamma^5\hat{\psi}(z,t) \\ -\hat{\psi}^\dagger(z,t)\gamma^5\hat{\psi}(z+\varepsilon,t) \end{bmatrix} | 0 \rangle = \frac{2i}{\pi\varepsilon} \tag{8.6}$$

which is equivalent to (5.13).

## References.